# On the Theoretical Analysis of Parametric Amplification of Femtosecond Laser Pulses in Crystals with a Regular Domain Structure

O. I. Sobirov[a], D. B. Yusupov[a], N. A. Akbarova[a], and U. K. Sapaev[a,*]

[a] Tashkent State Technical University named after I. Karimov, Tashkent, 100095 Uzbekistan

*e-mail: usapaev@gmail.com

**Abstract**—The parametric amplification of ultrashort (femtosecond) laser pulses in crystals with a regular domain structure (RDS) of the 5%Mg : PPLN type has been investigated theoretically. The focus was on the formation of a signal pulse in dependence of the dispersion of the medium and cubic nonlinearity. It is shown that a small deviation of the size of domains from their exact value, determined by the coherent length of nonlinear interaction of optical waves, may increase to a great extent the efficiency of signal-wave generation. The reason is that the phase shifts due to the third-order nonlinearity and dispersion of the medium (as a rule, they affect only the generalized phases of nonlinear wave interaction) may be partially compensated by the influence of the "excess" wave number of nonlinear lattice. The optimal domain sizes, at which the efficiency of signal-wave generation under conditions of self-action and nonstationarity becomes maximum, have been analyzed based on the results obtained.

## 1. INTRODUCTION

To solve most of applied and fundamental problems of laser physics, it is often necessary to convert laser frequencies into the spectral ranges in which direct lasing may be either impossible or low-efficient. To this end, various processes of frequency conversion in quadratic nonlinear optical media are used. Among them, parametric amplification (PA) of laser radiation takes a particular place. This process makes it possible to covert gradually laser frequency into various spectral ranges and amplify significantly the signal-wave energy in the presence of a high-power pump wave, provided that the phase-matching conditions in conventional anisotropic crystals (or quasi-phase-matching conditions in crystals with a regular domain structure (RDS)) are fulfilled.

PA can generally be implemented in bulk nonlinear optical crystals and periodically polarized crystals (see, e.g., [1, 2] and references therein). The latter approach is widely applied in current practice (due to the development of the growth technology for these crystals [3–5]), because these nonlinear optical lattices can be prepared for implementing PA with an arbitrary laser wavelength. Note that the quasi-phase-matching condition is satisfied with a change in the sign of second-order nonlinearity from domain to domain [6, 7].

The high energy efficiency of signal-wave generation under PA conditions is an important characteristic, and one should always try to implement it in practice. To this end, high-intensity pump radiation generated by femtosecond lasers must be used. However, the use of ultrashort (shorter than ~100 fs [7]) high-energy laser pulses may lead to the occurrence of two limiting effects: (i) an increase in the influence of the third-order nonlinearity and (ii) an increase in the influence of the medium dispersion. The former effect depends linearly on the incident radiation intensity and changes the refractive index of the medium, which violates the quasi-phase-matching condition. The latter effect leads to a group delay of interacting pulses because of their different group velocities and induces a dispersion spread because of the different phase velocities of different frequency (Fourier) components of the pulses. Both these effects limit significantly the efficiency of signal-wave generation at PA of ultrashort laser pulses.

Note that the aforementioned problem was solved successfully using a chirped signal-wave pulse at the input of a nonlinear crystal [8, 9]. However, this method requires complex experimental setups and may cause partial energy loss in both signal and pump waves.

The purpose of this study was to show theoretically that a small variation in the domain size in RDS crystals at a chosen PA process affects significantly the formation of a signal-wave pulse for ultrashort laser pulses. The reason is that the phase shifts caused by linear phase mismatch, cubic nonlinearity, and dispersion of the medium are partially compensated by the reciprocal-lattice vector of modulation of RDS crystal nonlinear susceptibility.

## 2. BASIC EQUATIONS OF PARAMETRIC AMPLIFICATION UNDER NONSTATIONARITY AND SELF-ACTION CONDITIONS

Under PA conditions, a high-frequency pump wave $\omega_p$ and a weak signal wave $\omega_s$ enter a nonlinear crystal. According to the condition $\omega_p = \omega_s + \omega_i$, the signal wave $\omega_s$ is amplified, and an idler wave $\omega_i$ is generated at the crystal output. The PA process under nonstationarity and self-action conditions is described by the following system of equations (ee–e interaction) [2, 10–12]:

$$\hat{M}_s A_s = -\frac{4\pi i d_{\text{eff}}}{n_s \lambda_s} \delta(z) A_i^* A_p e^{-i\Delta kz} - \frac{3\pi i}{n_s \lambda_s} A_s \chi^{(3)} \left[ |A_s|^2 + 2|A_i|^2 + 2|A_p|^2 \right],$$

$$\hat{M}_i A_i = -\frac{4\pi i d_{\text{eff}}}{n_i \lambda_i} \delta(z) A_s^* A_p e^{-i\Delta kz} - \frac{3\pi i}{n_i \lambda_i} A_i \chi^{(3)} \left[ |A_i|^2 + 2|A_s|^2 + 2|A_p|^2 \right], \quad (1)$$

$$\hat{M}_p A_p = -\frac{4\pi i d_{\text{eff}}}{n_p \lambda_p} \delta(z) A_i A_s e^{i\Delta kz} - \frac{3\pi i}{n_p \lambda_p} A_p \chi^{(3)} \left[ |A_p|^2 + 2|A_s|^2 + 2|A_i|^2 \right]$$

with the boundary conditions

$$A_p(z=0,t) = A_0 \exp\left[-2\ln 2(t/\tau)^2\right],$$
$$A_s(z=0,t) \approx 10^{-4} A_p(z=0,t), \tag{2}$$
$$A_i(z=0,t) = 0.$$

Here, $A_p$, $A_s$, and $A_i$ are the complex amplitudes of the pump, signal, and idler waves, respectively; $A_0$ is the maximum real pump amplitude at the crystal input; $\lambda_p$, $\lambda_s$, and $\lambda_i$ are the wavelengths of the pump, signal, and idler waves; $n_p$, $n_s$, and $n_i$ are the refractive indices for the pump, signal, and idler waves; $d_{\text{eff}} = \chi^{(2)}/2$ and $\chi^{(3)}$ are the nonlinear coupling coefficients of the second and third orders, respectively; $\delta(z)$ is an alternating periodic function; $\tau$ is the pump-pulse width at the input of nonlinear crystal (at the FWHM intensity level); $\Delta k = k_p - k_s - k_i$ is the phase mismatch; and $\hat{M}_j = \partial/\partial z + D_{1j}\partial/\partial t - (i/2)D_{2j}\partial^2/\partial t^2 + (1/6)D_{3j}\partial^3/\partial t^3$, where $D_{1j} = dk/d\omega|_{\omega=\omega_j}$, $D_{2j} = d^2k/d\omega^2|_{\omega=\omega_j}$, and $D_{3j} = d^3k/d\omega^3|_{\omega=\omega_j}$ ($j = \{i, p, s\}$).

## 3. ANALYTICAL SOLUTION OF THE PROBLEM

Before showing why the optimality of the PA process under nonstationarity and self-action conditions is not determined by the domain length (equal to the coherent length $\pi/|\Delta k|$), it is relevant to refer to the analytical solutions to system of equations (1).

System of equations (1) can be solved analytically using the given-field approximation and a moving coordinate system and restricting oneself to the first approximation of the dispersion theory (on the assumption that the pump radiation is a plane wave). Then (1) can be written in the spectral space as

$$\frac{\partial A_s}{\partial z} + i\omega(D_{1s} - D_{1i})A_s = -i\frac{8d_{\text{eff}}}{n_s\lambda_s}A_i^* A_0 e^{-i\delta kz} - i\frac{6\pi}{n_s\lambda_s}A_s\chi^{(3)}|A_0|^2,$$
$$\frac{\partial A_i}{\partial z} = -i\frac{8d_{\text{eff}}}{n_i\lambda_i}A_s^* A_0 e^{-i\delta kz} - i\frac{6\pi}{n_i\lambda_i}A_i\chi^{(3)}|A_0|^2, \tag{3}$$

where $\delta k = \Delta k - \pi/d$ ($d$ is the domain thickness). Equation (3) can be simplified by introducing new designations for the complex amplitudes of signal and idler waves, respectively:

$$A_s(z,\omega) = U_s(z,\omega) e^{-iz(\omega(D_{1s}-D_{1i})+\alpha_s)}, \qquad A_i(z,\omega) = U_i(z,\omega) e^{-i\alpha_i z}, \qquad (4)$$

where $\alpha_s = (6\pi/n_s\lambda_s)|A_0|^2 \chi^{(3)}$ and $\alpha_i = (6\pi/n_i\lambda_i)|A_0|^2 \chi^{(3)}$.

Having introduced (4) into (3), we obtain the well-known equations for PA within the specified-field approximation:

$$\frac{\partial U_s(z,\omega)}{\partial z} = -i\beta_s U_i^* e^{-i\Delta S(\omega)z}, \qquad \frac{\partial U_i(z,\omega)}{\partial z} = -i\beta_i U_s^* e^{-i\Delta S(\omega)z}, \qquad (5)$$

where $\beta_s = 8A_0 d_{\text{eff}}/n_s\lambda_s$, $\beta_i = 8A_0 d_{\text{eff}}/n_i\lambda_i$, and $\Delta S(\omega) = -\alpha_s - \omega(D_{1s} - D_{1i}) - \alpha_i - \delta k$. This system of equations has the following known solution for the signal wave [10, 13]:

$$U_s(z,\omega) = -U_{s0} \exp\left[-\frac{i\Delta S(\omega)z}{2}\right]\left[\text{ch}(bz) + i\left(\frac{\Delta S(\omega)}{2b}\right)\text{sh}(bz)\right], \qquad (6)$$

where $b(\omega) = (1/2)\sqrt{g^2 - \Delta S(\omega)^2}$ and $g = A_0^2 \beta_s \beta_i$.

One can easily see that the solution to Eq. (5) in the form (6) rises monotonically only at $\Delta S(\omega) \approx 0$. Hence, $\delta k \approx -\alpha_s - \omega(D_{1s} - D_{1i}) - \alpha_i$. One can find the optimal domain thickness:

$$d = \pi/\left[\Delta k - \alpha_s - \omega(D_{1s} - D_{1i}) - \alpha_i\right]. \qquad (7)$$

Having disregarded the Kerr-type nonlinearity and dispersion of the medium, one can find from (7) that $d = d_0 = \pi/|\Delta k|$. Using the real experimental condition (see below) in (7), we obtain $d \approx 0.87 d_0$.

The system of partial differential equations (1) with boundary conditions (2) is solved numerically using the split-step method [14]. The accuracy of numerical calculations was verified using the energy conservation law for the interacting waves and the value of quasi-static interaction length [10].

To date, femtosecond lasers operating in the visible and near-IR ranges have been developed well. However, in view of the importance of generation of ultrashort laser pulses in the mid-and far-IR spectral ranges, the use of specifically PA of ultrashort laser pulses in nonlinear optical media is of great interest [16, 17]. Therefore, a numerical analysis of the PA of ultrashort laser pulses was performed with the following real experimental parameters: a nonlinear photonic 5%Mg : PPLN crystal (its dispersion ratio and values of quadratic and cubic nonlinearities were reported in [15] and [12], respectively), $I_p = 10^{15}$ W/m$^2$, $I_s = 10^{-4} I_p$, $\lambda_p = 1.065$ μm (femtosecond laser pulses at this wavelength are generated by an ytterbium-doped fiber laser (see, e.g., [16])), $\lambda_s = 3.4$ μm (femtosecond pulses at this wavelength are generated by Ti : sapphire lasers using cascade PA or the difference-frequency generation process in nonlinear crystals (see, e.g., [17])), $\lambda_i \approx 1.5$ μm, and $d_0 \approx 15.3$ μm.

The role of dispersion at a chosen process of converting femtosecond pulse frequency can easily be estimated for the 5%Mg : PPLN crystal. To this end, we calculated the parameters $D_{1s} \approx 7.341$ fs/μm, $D_{2s} \approx -0.812$ fs$^2$/μm, $D_{3s} \approx -3.2513$ fs$^3$/μm, $D_{1i} \approx 7.248$ fs/μm, $D_{3i} \approx 0.812$ fs$^3$/μm, $D_{1p} \approx 7.342$ fs/μm, $D_{2p} \approx 0.234$ fs$^2$/μm, and $D_{3p} \approx 1.016$ fs$^3$/μm based on the corresponding dispersion relation [15]. Using these data, one can decide if it is necessary to take into account these parameters in the frequency conversion process under study.

Let us consider a laser pulse of width τ, incident on a nonlinear crystal with a length $L$. Then it is necessary to take into account the influence of the group-velocity mismatch if $L \geq L_v$; the group-velocity dispersion in the first approximation if $L \geq L_{dis1}$; and, finally, the group-velocity dispersion in the second approximation if $L \geq L_{dis2}$ [10]. Under the PA conditions these lengths are determined as $L_v = \tau/(D_{1s} - D_{1p})$, $L_{dis1} = 2\tau^2/D_{2s}$, and $L_{dis2} = 6\tau^3/D_{3s}$.

Figure 1 shows how $L_v$, $L_{dis1}$, and $L_{dis2}$ (dotted, dashed, and solid lines, respectively) change with a change in the incident pulse width $\tau$ for the 5%Mg : PPLN crystal at $\lambda_s = 3.4$ µm and $\lambda_p = 1.065$ µm.

Based on Fig. 1, one can determine the effect that must be taken into account in dependence of the periodic-crystal length and the width of interacting pulses when calculating the PA of femtosecond pulses. Let us have a periodic crystal 1 cm long and a pulse with a width $\tau \gtrsim 75$ fs. Then this problem can be considered as time-independent. Beginning with $\tau \lesssim 75$ fs, one must take into account the group-velocity dispersion in the first approximation (i.e., the effect of pulse dispersion spreading) in the calculations. At the same time, if $\tau \lesssim 20$ fs, one must take into account all three approximation orders of the dispersion theory. Note that all three approximation orders were taken into consideration in our numerical calculations (however, depending on the chosen crystal length and the width of incident pulses (signal and pump waves), the influence of a particular term on the formation of signal-wave pulse may be negligible).

## 4. RESULTS AND DISCUSSION

The results for a 10-fs pump pulse are presented in Fig. 2. The solid curve shows the signal-wave efficiency obtained with allowance for only the third-order nonlinearity. The maximum efficiency is obtained for a periodic crystal approximately 0.5 mm thick (solid curve). Note that the results barely changed with all limiting factors disregarded (this case is omitted). However, the signal-wave generation efficiency (dashed curve) decreases sharply when the dispersion of the medium is taken into account. With all limiting factors taken into consideration, the signal-wave generation efficiency decreases even more strongly (the dotted curve). The results obtained show that the main limiting factor for the signal-wave generation efficiency in these ranges of the pump pulse width and intensity is the dispersion of the medium rather than the third-order nonlinearity (see also [11] and Fig. 1).

The role of dispersion in the formation of signal-wave pulse can clearly be seen in Fig. 3, which also shows the dependence of the signal-wave efficiency on the thickness of domains in RDS crystals at different $\tau$ values. One can see in Fig. 3 that the dispersion limits strongly the signal-wave generation efficiency only for pump pulses shorter than ~10 fs.

Then, we investigated the influence of the domain thickness on the efficiency of signal-wave generation. Figure 4 shows the dependence of the signal-wave generation efficiency on the domain thickness at a fixed value $z = 0.6$ cm (i.e., the total length of RDS crystal does not change with a change in the domain thickness) and the width $\tau = 10$ fs of the pump and signal wave pulses. One can see that the efficiency of signal-wave generation reaches a maximum at $d/d_0 \approx 0.7955$ (as was expected proceeding from the analytical results) rather than at $d/d_0 = 1$. Here, $d_0$ is the thickness of a single domain, which is equal to the coherent length of the given nonlinear optical process of frequency conversion. Therefore, we can conclude that the phase shifts caused by limiting factors are compensated for with a decrease in the domain thickness. Note that the gain in the efficiency of signal-wave generation is approximately twice higher.

Then, using the results presented in Fig. 4, we repeated calculations to determine the dynamics of energy exchange between the interacting waves at $d/d_0 = 1$ and $d/d_0 \approx 0.7955$. Figure 5 shows the dependence of the signal-wave generation efficiency on $z$ at $d/d_0 = 1$ (dotted curve) and $d/d_0 \approx 0.7955$ (dashed curve). Here, one can clearly see the gain in the signal-wave generation efficiency with an increase in the length of nonlinear interaction between the waves. The results for the width $\tau = 5$ fs of the pump and signal wave pulses and $d/d_0 \approx 0.7955$ are also shown for comparison.

One can also see in Fig. 5 that the dependence of generation efficiency on $z$ at $d/d_0 \approx 0.7955$ is oscillating. The reason is that a small deviation of domain sizes from the average value leads to an additional phase mismatch, which can be partially compensated by additional phase changes, caused by cubic nonlinearity and frequency dispersion of the medium. However, complete compensation is impossible because the formulas describing these changes are significantly different. This fact explains the oscillating dependence of the efficiency of signal-wave generation on the interaction length (see Fig. 5).

## 5. CONCLUSIONS

The parametric amplification of ultrashort (femtosecond) laser pulses in 5%MgO:LiNbO$_3$ crystals with a regular domain structure was investigated numerically. The influence of cubic nonlinearity and dispersion effects on the formation of a signal-wave pulse using bandwidth-limited femtosecond pulses was analyzed. The numerical calculations were performed for bandwidth-limited pump pulses and signal waves with widths of ~100, ~50, ~10, and ~5 fs. It was shown that the domains have an optimal length (differing from $d_0 = \pi/\Delta k$), at which the efficiency of the frequency-conversion process under consideration (signal-wave generation) increases sharply. Calculations showed that a small deviation of the domain size from the exact value, which provides the quasi-phase-matching condition, makes it possible to compensate partially the phase shift caused by the influence of cubic nonlinearity and dispersion of the medium. The technique in use can be applied to optimize other frequency-conversion processes if the influence of cubic nonlinearity and dispersion of the medium becomes significant.

## FUNDING

This study was supported in part by project nos. Uzb-Ind-2021-96, Uzb-Ind-2021-83, and ATLANTIC-823897 (HORIZON-2020).

CONFLICT OF INTEREST

The authors declare that they have no conflicts of interest.

REFERENCES

FIGURE CAPTIONS

**Fig. 1.** (Color online) Dependences of the (dotted line) $L_v$, (dashed line) $L_{dis1}$, and (solid line) $L_{dis2}$ values on the width τ of the pulses incident on an RDS crystal (5% Mg : PPLN); $λ_s$ = 3.4 µm, $λ_p$ = 1.065 µm. The same dependences for ultrashort pulses are shown in the inset.

Key:

фс → fs;

см → cm.

**Fig. 2.** Dependence of the energy efficiency of signal-wave generation on the thickness of nonlinear photonic crystal. The results are presented for the width τ = 10 fs of the pump and signal pulses, taking into account the influence of (solid line) only cubic nonlinearity, (dashed line) only dispersion of the medium, and (dotted line) all limiting factors ($d = d_0$).

**Fig. 3.** Dependence of the signal-wave generation efficiency on the thickness of the nonlinear photonic crystal for the following widths of pump and signal pulses: τ = (solid line) 100 fs, (dashed line) 50 fs, and (dotted line) 10 fs ($d = d_0$).

**Fig. 4.** Dependence of the signal-wave generation efficiency on $d = d_0$ at $z = 0.6$ cm and the width τ = 10 fs of pump and signal pulses.

**Fig. 5.** Results of numerical calculation of signal-wave generation efficiency with allowance for all limiting factors, at the following widths of the pump and signal

pulses: (dotted and dashed lines) $\tau = 10$ fs ($d/d_0 = 1$ and ~0.7955, respectively) and (solid line) $\tau = 5$ fs ($d/d_0 \approx 0.7955$).